\documentclass[english,letterpaper,american,conference]{IEEEtran}
\usepackage[T1]{fontenc}
\usepackage[latin1]{inputenc}
\usepackage{graphicx}

\newtheorem{example}{Example}
\newtheorem{lemma}{Lemma}
\newcommand{\lyxrightaddress}[1]{
\par {\raggedleft \begin{tabular}{l}\ignorespaces
#1
\end{tabular}
\vspace{1.4em}
\par}
}

\makeatletter

\usepackage[draft]{hyperref}

\usepackage{times}

\makeatother
\IEEEoverridecommandlockouts
\usepackage{babel}

\begin{document}

\title{Construction of Codes for Network Coding}

\author{Andreas-Stephan Elsenhans, Axel Kohnert, Alfred Wassermann }
\maketitle
\begin{abstract}
Based on ideas of Kötter and Kschischang \cite{koetter-2007} we use
constant dimension subspaces as codewords in a network. We show a
connection to the theory of q-analogues of a combinatorial designs,
which has been studied in \cite{Braun_kerber_laue_q_analoga} as a
purely combinatorial object. For the construction of network codes
we successfully modified methods (construction with prescribed automorphisms)
originally developed for the q-analogues of a combinatorial designs.
We then give a special case of that method which allows the construction
of network codes with a very large ambient space and we also show
how to decode such codes with a very small number of operations.
\end{abstract}
network coding, q-analogue of Steiner systems, subspace codes, constant
dimension subspace codes, decoding, Singer cycle

\section{Introduction}

\subsection{Subspace Codes}

In \cite{koetter-2007} R. Kötter and F. R. Kschischang developed
the theory of subspace codes for applications in network coding. We
will modify their presentation in some way. We denote by $L(GF(q)^{v})$
the lattice of all subspaces of the \emph{ambient} space, which is
a vector space of dimension $v$ over the finite field with $q$ elements.
The partial order of $L(GF(q)^{v})$ is given by inclusion. A \emph{subspace
code} $C$ then is a subset of $L(GF(q)^{v})$. A \emph{constant dimension
code} is the special case where all subspaces in $C$ are of the same
dimension. To study error correcting codes we have to define some
distance between codewords (in this case codewords are subspaces).
The most natural one is the graph theoretic distance in the \emph{Hasse
diagram} (vertices are the elements of $L(GF(q)^{v})$ and two subspaces
are connected by an edge if they are direct neighbors in the partial
order) of the lattice $L(GF(q)^{v})$. An equivalent definition without
using the underlying graph is as follows:

The \emph{subspace distance} between two spaces $V$ and $W$ in $L(GF(q)^{v})$
is defined as\[
d_{S}(V,W):=\dim(V+W)-\dim(V\cap W)\]
 which is equal to\[
\dim(V)+\dim(W)-2\dim(V\cap W).\]

This defines a metric on $L(GF(q)^{v})$. Like in classical coding
theory we define the \emph{minimum (subspace) distance} of a subspace
code $C$.\[
D_{S}(C):=\min\{d_{S}(V,W):V,W\in C\mbox{\ensuremath{\,\,} and }V\not=W\}.\]

We can now define the optimal (subspace) code problem:
\begin{quotation}
(P) For fixed parameters $q,v,d$ we want to find the maximal number
$m$ of subspaces $V_{1},\ldots,V_{m}$ in $L(GF(q)^{v})$ such that
the corresponding subspace code $C=\{V_{1},\ldots,V_{m}\}$ has at
least minimum distance $d.$ 
\end{quotation}
This is a specific instance of a packing problem in a graph. In classical
coding theory the underlying graph is the Hamming graph. For network
codes it is the Hasse diagram of the linear lattice. In the case of
a binary code the Hamming Graph is isomorphic to the Hasse diagram
of the powerset lattice. Using this connection we can look at the
optimal (subspace) code problem as the \emph{$q-$analogue} of the
classical optimal code problem in the Hamming graph. To get the $q-$analogue
we have to substitute a subset (= $0/1$ sequence of length $v$)
of size $k$ by a $k-$dimensional subspace of $GF(q)^{v}.$ The number
of $k-$dimensional subspaces of $GF(q)^{v}$ is denoted by the Gaussian
coefficient $\left[\begin{array}{c}
v\\
k\end{array}\right]_{q}$.

\subsection{$q-$Analogues of Designs\label{sub:Analogues-of-Designs}}

A \emph{$t-(v,k,\lambda)$ design} is a set $C$ of $k-$element subsets
(called blocks) of the set $\{1,\ldots,v\}$ such that each $t-$element
subset of $\{1,\ldots,v\}$ appears in exactly $\lambda$ blocks.
The special case of $\lambda=1$ is called a \emph{Steiner system}.
Like in the subsection above we now define the $q-$analogue of a
$t$-design. A $t-(v,k,\lambda)$ design over the finite field $GF(q)$
is a multiset $C$ of $k-$dimensional subspaces (called $q$-blocks)
of the $v$-dimensional vector space $GF(q)^{v}$ such that each $t-$dimensional
subspace of $GF(q)^{v}$ is a subspace of exactly $\lambda$ $q-$blocks.
The connection with the constant dimension codes is given by the following
observation in the case of a Steiner system: Given a $q-$analogue
of a $t-(v,k,1)$ design $C$ we get a constant dimension code of
minimum distance $2(k-t+1)$, since each $t$-dimensional space is
contained in exactly one $k$-dimensional subspace the intersection
between two spaces from $C$ is at most $(t-1)-$dimensional. Therefore
the minimum distance of $C$ is at least $2(k-t+1).$ On the other
hand given any $(t-1)-$dimensional subspace $V$ we can find two
$t-$dimensional spaces $U,W$ with intersection $V$ and then two
unique $q-$blocks containing $U$ and $W.$ The minimum distance
between these $q-$blocks is $2(k-t+1).$ $q$-analogues of designs
were introduced by Thomas in 1987 \cite{thomas_q_analoga_1987}. Later
they were studied in a paper by Braun et al. \cite{Braun_kerber_laue_q_analoga}
where the authors constructed the first non-trivial $q-$analogue
of a $3$-design. We will describe a method in the second section
which is based  on their paper and which we use to construct constant
dimension codes. First results found using that method were already
in \cite{pre05496601}.

\subsection{Encoding/Decoding}

In \cite{koetter-2007} the authors introduced the operator channel
as a model to study subspace codes. The input and output alphabet
of the channel is the lattice $L(GF(q)^{v})$. The transmission of
an information coded by a space $U$ works as follows: The transmitter
inserts into the network vectors from $U.$ During transmission the
internal nodes of the network receive several such vectors and forward
an arbitrary linear combination of the received vectors. The receiver
collects the incoming vectors and tries to rebuild $U$ as the space
generated by the incoming vectors. There are two possible problems
at the receiver. There can be \emph{erasures}, which means that some
vectors are missing and the generated space is a subspace of $U.$
Or there can be \emph{errors}, which means that the receiver got vectors
not from $U.$ In the third section we will give a decoding algorithm
for minimum distance decoding which allows error and erasure correction
for a special class of constant dimension codes. This new method uses
the symmetries of this special class of network codes, which were
also studied by Etzion and Vardy \cite{2010arXiv1004.1503E}, who
called these codes cyclic.

\section{Construction of Constant Dimension Subspace Codes}

In the following we work only with constant dimension codes built
by a collection of $k-$dimensional subspaces of $GF(q)^{v}$ of minimum
distance at least $2(k-t+1)$ as described in section \ref{sub:Analogues-of-Designs}.
In \cite{pre05496601} we gave a general method using a prescribed
group $G$ of automorphisms of a putative constant dimension code
$C$ to give an equivalence between the existence of such a code and
a solution of a Diophantine system of inequalities. Therefore the
construction of such a code $C$ boils down to finding a $(0/1)-$solution
$x$ of a Diophantine system of inequalities of the form:\[
M^{G}x\le\left(\begin{array}{c}
1\\
\vdots\\
1\end{array}\right).\]
The number of rows in $M^{G}$ is the number of orbits of $G$ on
the $t-$dimensional subspaces of $GF(q)^{v}$ and the number of columns
is the number of orbits of $G$ on the $k-$dimensional subspaces
of $GF(q)^{v}$. In \cite{pre05496601} we described the general method
for the construction using arbitrary automorphisms, and a further
variant which is useful in the case of $G$ equal to the group generated
by a Singer cycle $S$. Orbits of the Singer cycle has been studied
for several applications \cite{0998.20003}. The nice property of
the Singer cycle is that $G=\left\langle S\right\rangle $ acts transitively
on the one-dimensional subspaces of $GF(q)^{v}.$ So we can label
any one-dimensional subspace $W$ by the unique exponent $i$ between
$0$ and $l:=\left[\begin{array}{c}
v\\
1\end{array}\right]_{q}-1$ with the property that $W=g^{i}V,$ where $V$ is some arbitrary
one-dimensional subspace. Given a $k$-space $U$ (for $1\le k\le v$)
we can describe it by the set $P_{U}$ of one-dimensional (i.e. numbers
between $0$ and $l$) subspaces contained in $U$. Given such a description
of a $k$-space $U$ it is now easy to get all the spaces building
the orbit under the Singer subgroup $G$. Take the set $P_{U}$ of
$\left[\begin{array}{c}
k\\
1\end{array}\right]_{q}$ numbers from $\{0,\ldots,l\}$ representing the one-dimensional subspaces
of $GF(q)^{v}$ being also subspaces of $U$ and now the action of
$S$ on a number in $P_{U}$ is simply adding one modulo $l.$ If
we look at $GF(q)^{v}$ as an field extension of degree $v$ of the
base field $GF(q)$ this is the use of a primitive element $\omega$
and writing the elements of $GF(q)$ using $\omega$ and an exponent.
\begin{example}
We study the case $q=2,v=5,k=2:$ A two-dimensional binary subspace
contains three one-dimensional subspaces. We get a two-dimensional
space by taking the two one-dimensional spaces labeled $\{0,1\}$
and the third one given by the linear combination of these two will
have a certain number, in this example $\{14\}$. Therefore we have
a two dimensional space described by the three numbers $\{0,1,14\}$.
To get the complete orbit under the Singer subgroup we simply have
to increase the numbers by one for each multiplication by a generator
$S$ of the Singer subgroup. The orbit length of the Singer subgroup
is $31$ and the orbit is built by the $31$ sets: $\{0,1,14\},\{1,2,15\},\ldots,\{16,17,30\},\{0,17,18\},\ldots$
$\{12,29,30\},$ $\{0,13,30\}$. 
\end{example}

\section{Real Coding}

In this chapter we will restrict to the case of a constant dimension
subspace code built from a single orbit $O$ of the Singer cycle on
the $3-$dimensional subspaces of $GF(2)^{v}$ with the property that
two subspaces of the orbit intersect in a subspace of dimension less
or equal to one. We will call such an orbit a \emph{good} orbit. The
subspaces in a good orbit $O$ form a code of minimum distance at
least $4.$ This is a less restrictive version of the codes studied
in \cite{DBLP:journals/corr/abs-0805-0507} where a code was studied,
where the subspaces intersect zero-dimensional.

We describe a single subspace $U$ in the orbit $O$ by the $7$ one-dimensional
subspaces contained in $U.$ As an example lets assume $U=\left\{ 0,1,4,10,18,23,25\right\} $.
The orbit-type can be described by the two distance between $v_{1},v_{2}\in U,$
i.e. the two exponents $i_{1},i_{2}$ of the generator $S$ defined
by $S^{i_{1}}v_{1}=v_{2}$ and $S^{i_{2}}v_{2}=v_{1}$. In this example
it is visualized by the complete graph $K_{V}$ with vertices from
$V$ and edge-labels given by the the smaller distance. For this concrete
example we get:

\begin{center}
\includegraphics[scale=0.4]{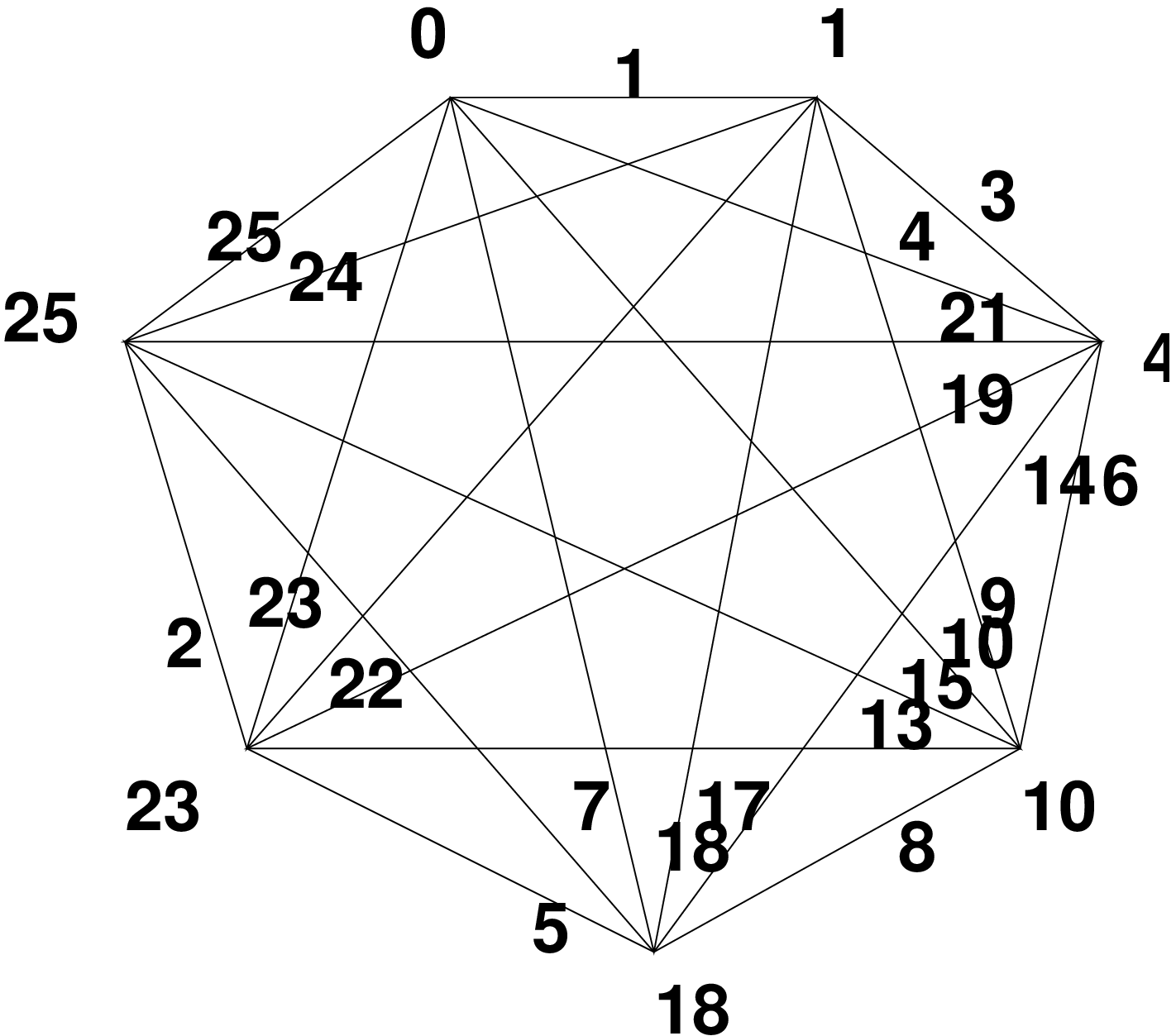}
\par\end{center}

If we take a different subspace from $O$ the vertex labels are increased
by a fixed number, but the pairwise distances remain unchanged. We
denote by $K_{O}$ the complete graph with $7$ points without vertex-labels
and edge-labels given by any $V\in O$. The following observation
is crucial for the proposed algorithm for decoding:
\begin{lemma}
All edge-labels in $K_{O}$ are different.
\end{lemma}
This is because otherwise we would have no orbit where two subspaces
intersect at most one-dimensional. This property is similar to the
questions studied in the theory of finite difference sets or Golomb
rulers (e.g. \cite{HandbookCombinatorialDesigns2_2007} p.419ff).
Above labeling comes from an optimal Golomb ruler. It is not clear
whether there is a primitive element in $GF(2^{v})$ such that we
get a vector space using above exponents. 

If we work with the field $GF(2^{v})$ instead of the vector space
$GF(2)^{v}$, than above graph is constructed using the elements $\{v_{1},\ldots,v_{7}\}$
of a subspace as labels of the vertices and the two quotients $v_{i}/v_{j}$
and $v_{j}/v_{i}$ as labels of the edges.

\subsection{Decoding}

There are two cases, which have to be checked for decoding.

\subsubsection{Erasure Case}

The first one is an erasure, meaning that we receive only a two-dimensional
subspace $U$, represented by two of the three one-dimensional subspaces
$U=\{r,s\}.$ The idea now is to identify the two vertices in $K_{O}$
corresponding to these two elements $r$ and $s.$ To do this we only
 have to compute the quotient (in the field $GF(2^{v})$ ) of the
two field elements $r$ and $s.$ And using the result we can lookup
the corresponding edge in $K_{O}$ and get vertex labels by taking
$r$ and $s.$ After that we use one further application of some power
of the Singer cycle (i.e. multiplication in $GF(2^{v})$) to get a
third independent vector, which finished the reconstruction of the
transmitted vectorspace. 

Therefore decoding including error correction in the erasure case
is one division and one multiplication in $GF(2^{v})$.

\subsubsection{Error Case}

The second case is that we received a vector not in the transmitted
space $V.$ As we have minimum distance $4$ the erroneous codeword
we want to correct is a $4-$dimensional space $U$ containing the
transmitted codeword $V.$ Again the idea is to identify the 3-dimensional
subspace $V$ inside $U$ by looking at the pairwise distances. We
start with $3$ independent vectors $r,s,t$ in $U.$ We know that
the space $I$ generated by $\{r,s,t\}$ intersects with $V$ in a
at least two dimensional space $W.$ Now we loop over all $7$ 2-dimensional
subspaces of $I$ and for each such space we try the reconstruction
like in the erasure case. We additionaly check whether the constructed
third vector is in the received word $U.$ If this is the case we
have found the original codeword $V.$

 Therefore decoding including error correction in the error case
needs at most $7$ divisions and $7$ multiplications in $GF(2^{v})$.

\subsection{Generalisations}

There are several immediate generalisations. Of course one can use
$n$ different orbits of the Singer cycle with the pairwise intersection
property for all spaces in the $n$ orbits. In this case you would
have $n\cdot21$ possible pairs of distances (=quotients). One can
also use $k-$dimensional subspaces instead of $3-$dimensional subspaces.
In this case one has to look at $\left(2^{k}-1\right)\left(2^{k}-2\right)/2$
pairs of distances and would get a code which allows to correct $k-2$
errors. In the non-erasure case error correction becomes more difficult
as one has to study more two-dimensional subspaces.

\section{How to find a code}

In this section we give some arguments showing, that it is 'easy'
to find \emph{good} orbits. We study the problem for arbitrary $k$
(not only $k=3)$ and for arbitrary finite fields (not only $q=2$).
For the computation of an estimation how 'easy' it is we use a primitive
element $\omega.$ For a 'real' application we represent the field
elements by polynomials, as the internal nodes of the network have
to compute linear combinations, meaning doing exor operations in the
case of $q=2.$

We need to find one \emph{good} orbit $O$ of the Singer cycle such
that all $k$-dimensional (in the above special case we had $k=3$)
subspaces  intersect pairwise at most in a one-dimensional subspace.
To do so, we choose arbitrarily $k$ representatives $a_{1},a_{2},\ldots,a_{k}$
of one-dimensional subspaces generating a $k-$space $U$ represented
by the one-dimensional subspace $P_{U}=$ $\left\{ b_{0},\ldots,b_{l}\right\} $(with
$l=\left[\begin{array}{c}
k\\
1\end{array}\right]_{q}-1$). If the ${l+1 \choose 2}$ quotient pairs built form $b_{i},b_{j}$
for $i>j$ are all different (meaning that the subspaces generated
from the quotient are different),  the intersection of all elemens
of the orbit of this space under the Singer cycle is at most one-dimensional.
Thus, the elements of the orbit $O$ form a code with minimum distance
$2(k-1)$. In order to generate such a orbit, we first select a primitive
element $\omega$. Then we choose randomly the $k$ representatives
of one-dimensional subspaces $a_{1},a_{2},\ldots,a_{k}$ by choosing
$k$ random numbers $e_{i}$ between $0$ and $q^{v}-2$ and setting
$a_{i}=\omega^{e_{i}}$, $i=1,\ldots,k$. The subspace $\langle a_{1},a_{2},\ldots,a_{k}\rangle$
will be the generator of the orbit $O$. The probability that the
dimension of $\langle a_{1},a_{2},\ldots,a_{k}\rangle$ is less than
$k$ is very small. In practice, for $k=3$ and $q=2$ we choose $e_{1}=0$,
$e_{2}=1$ and a random number $1<e_{3}\leq2^{v}-2$.

The property that all pairwise differences (mod $q^{v}-1$) between
all $\left[\begin{array}{c}
k\\
1\end{array}\right]_{q}$ non-zero elements of the subspace generated by $a_{1},a_{2},\ldots,a_{k}$
are different seems to be randomly distributed. For example, for two
generators $a_{i}$ and $a_{j}$ the integer number $0\leq x\leq q^{v}-2$
with $\omega^{x}=a_{i}+a_{j}$ is likely to be hard to compute. It
is called the discrete logarithm problem in cryptography. This problem
is also closely related to the problem of computing the Jacobi logarithm
\cite{030.0860cj}, also called Zech's logarithm. There are no efficient
(polynomial time) algorithms known to compute the discrete logarithm
\cite{343482}, also the discrete logarithm of the sum of arbitrary
elements with known discrete logarithm as in our case seems to be
without structure.

The overall number $m$ of possible pairs $\{b_{i},b_{j}\}$ (or pairs
$\{e_{i}-e_{j}\bmod q^{v}-1,e_{j}-e_{i}\bmod q^{v}-1\}$) is equal
to $(q^{v}-2)/2$ for $q$ even, and $\left(q^{v}-1\right)/2$ for
$q$ odd. Let $s$ be the number of (unordered) pairs of one-dimensional
subspaces in a $k$-dimensional vector space. The probability that
all $s$ pairs $\{b_{i},b_{j}\}$ are different is equal to \[
(1-\frac{1}{m})(1-\frac{2}{m})\cdots(1-\frac{s}{m})\]
 which is approximately \[
\biggl(1-\frac{s}{2m}\biggr)^{s-1}\approx e^{-s(s-1)/2m}.\]
 For example, for $q=2$, $v=100$ and $k=3$ we get $s={2^{3}-1 \choose 2}=21$,
$m=633825300114114700748351602687$ and therefore \[
-s(s-1)/2m\approx-3.3132158019282496\cdot10^{-28}.\]
 So, it is extremely unlikely to find a random orbit which does not
fulfill the desired property.

If we take the union of $n$ orbits as our code we can apply the same
estimation. For $q=2$ we get \[
s={n\cdot(2^{k}-1) \choose 2}.\]
 In the example above, the expected number of orbits which can be
combined without conflicts is $66\,955\,225\,653\,132$. 

~

~\\
~\\
~\\
\\
\\
\\
\\
\\
\\
\\
\\
\\
\\
~\\
~\\
~\\
~\\
~\\
~\\
~

\bibliographystyle{plain}
\bibliography{codes,incidence}

\lyxrightaddress{Department of Mathematics\\
University of Bayreuth\\
 95440 Bayreuth\\
 Germany\\
 \{stephan.elsenhans,axel.kohnert,alfred.wassermann\}\\
@uni-bayreuth.de\\
\today}
\end{document}